# Design and empirical validation of a stock-Android software architecture for Wi-Fi Direct multi-group communication

Kwasi Edwards[a,1], Wayne Goodridge[a], Koffka Khan[a], Amit Ramkissoon[a]

[a]*Department of Computing and Information Technology, The University of the West Indies, St. Augustine, Trinidad and Tobago*

Abstract

Context: Stock Android exposes Wi-Fi Direct peer-to-peer APIs, but it does not provide application-transparent communication across multiple WiFi Direct groups. For developers working on non-rooted devices, the main obstacle is architectural: interface-specific transport contexts, relay roles, and forwarding state must be coordinated entirely at application level.

Objectives: This paper investigates whether multi-group Wi-Fi Direct communication can be realized as a stock-Android software architecture while preserving forwarding-state consistency and remaining compatible with Android 11 devices without rooting or operating-system modification.

Methods: We design SWARNET, a layered artifact composed of a Flutter application layer, a Kotlin native networking layer, interface-bound P2P and legacy-Wi-Fi sockets, relay-state management, and subscription-based forwarding tables. We evaluate the implemented artifact on five stock Samsung Galaxy A10s smartphones across four single-group and multi-group scenarios using archived throughput and packet-loss measurements.

Results: The artifact remained operational in all four scenarios. Peak receiver throughput observed from the archived curves was approximately 19.7 Mbit/s in 2d1g, 17.9 Mbit/s in 3d1g, 16.1 Mbit/s in 4d2g, and 16.0 Mbit/s in 5d3g. Packet

---

[1] Corresponding author

*Email addresses:* kwasi.edwards@sta.uwi.edu (Kwasi Edwards), wayne.goodridge@sta.uwi.edu (Wayne Goodridge), koffka.khan@sta.uwi.edu (Koffka Khan), amit.ramkissoon@sta.uwi.edu (Amit Ramkissoon)

loss increased with forwarding complexity, reaching about 19–20% only in the highest-load region of the three-group case.

Conclusion: The contribution is an implementable software architecture and a feasibility study showing that stock-Android multi-group Wi-Fi Direct communication can be engineered at application level on non-rooted devices. The results support architectural feasibility in a small static testbed; they do not establish broad resilience, scalability, or deployment readiness.

1. Introduction

Wi-Fi Direct (WFD) enables device-to-device communication over commodity smartphone hardware without requiring an infrastructure access point. On Android, the Wi-Fi P2P framework supports peer discovery, connection establishment, and group management through the WifiP2pManager APIs [1, 2]. In each WFD group, one device acts as the Group Owner (GO) and the remaining devices join as clients. This model is adequate for singlegroup communication, but it does not provide application-transparent routing across multiple WFD groups.

For practitioners, the difficulty is not merely a networking limitation. It is also a software-engineering problem arising from how stock Android constrains application behavior. Applications running on non-rooted devices cannot alter interface addressing or install custom network-layer forwarding rules. In addition, each WFD GO uses the same default P2P address (192.168.49.1), which prevents straightforward GO-to-GO forwarding across groups [5, 4]. The developer is therefore forced to solve multi-group communication at application level, while also dealing with API changes introduced in modern Android versions, interface binding through multinetwork APIs, state consistency across relays, and partial automation of credentials and group transitions.

Earlier studies established that multi-group operation is feasible by means of relay switching, simultaneous legacy/WFD connections, or applicationlayer tunnels [5, 4, 3, 6]. However, most of that literature primarily emphasizes network throughput or gateway behavior. The present paper instead focuses on the architecture of a software artifact that must operate under stock-Android constraints. The central research question is therefore: *how can a stock-Android* *software architecture realize the architectural feasibility of multi-group WFD communication while preserving forwarding-state consistency and remaining compatible with non-rooted devices?*

This reframing changes the contribution claimed by the paper. We do not present a new general routing theory or a formal optimization framework. Rather, we report the design and empirical validation of SWARNET, a stockAndroid software architecture that combines: (i) interface-aware transport binding; (ii) relay-state management; (iii) subscription-based forwarding consistency; and (iv) application-level control procedures for group join, relay promotion, publication forwarding, and disconnect recovery. The main contributions are as follows.

1. We define a stock-Android software architecture for WFD multi-groupcommunication that separates application logic, native Wi-Fi Direct control, relay coordination, and subscription-state management.

2. We document the architectural decisions required to support multi-groupcommunication on non-rooted Android 11 devices, including dual-interface socket binding and relay-centered forwarding.

3. We specify a consistency-oriented forwarding model in which subscriptionrouting tables are updated during join, subscribe, unsubscribe, relaypromotion, and disconnect events.

4. We provide an empirical feasibility study on Android 11 smartphones andinterpret the observed throughput and packet-loss curves as implementationlevel validation of the architecture.

The remainder of the paper is organized as follows. Section 2 reviews the problem context and positions the work relative to prior literature. Section 3 derives the architecture requirements imposed by stock Android. Section 4 presents SWARNET and its implementation decisions. Section 5 describes the artifact and empirical-validation methodology. Section 6 reports the results. Section 7 discusses portability, maintainability, and validity threats. Section 8 concludes the paper.

2. Problem context and related work

*2.1. Why stock-Android multi-group communication is a software-architectureproblem*

In single-group WFD operation, a GO provides access-point-like behavior for its clients. In multi-group operation, this model breaks down because each GO exposes the same default P2P address. If one GO tries to communicate with another GO that is simultaneously acting as a legacy client of an adjacent group, loopback and source-address conflicts arise [5, 4]. On rooted or modified platforms, a developer might address this through interface reconfiguration or custom routing rules. On stock Android, those options are unavailable.

The practical solution space is therefore defined by software-level choices: which process creates sockets on which network context, how interface affinity is preserved, which node acts as the forwarding anchor for inter-group traffic, how subscription state is propagated, and how the system recovers from relay churn without privileged APIs. That makes the design problem architectural rather than purely algorithmic.

*2.2. Closest prior studies*

Di Felice *et al.* used a time-sharing relay that periodically switched group affiliation to disseminate emergency messages across smartphone groups [3]. This design is useful in public-safety settings, but it introduces switching delay and relay buffering by construction. Funai *et al.* compared time sharing, UDP multicast, and hybrid simultaneous-connection methods on Android 4.4.2 and showed that simultaneous connections clearly outperform time sharing [4]. Casetti *et al.* proposed one of the earliest content-centric routing architectures over WFD multi-group topologies and demonstrated that the communication backbone is strongly affected by traffic direction and that GO-originated broadcast transmissions form a major bottleneck [5]. Teófilo *et al.* showed group-to-group bidirectional communication with relay nodes, further illustrating the importance of relay construction in WFD multi-group systems [6].

Although these studies are highly relevant, they are mostly framed as networking contributions. The gap addressed in the present paper is narrower: the literature offers limited guidance on how to package these ideas into a maintainable stock-Android software artifact that operates on modern Android versions without rooting.

## 2.3. Paper positioning

Table 1 summarizes how the present paper differs from the closest literature. The distinguishing point is not simply that SWARNET uses relays. It is that the implementation is organized as a stock-Android architecture with explicit attention to interface binding, relay-state roles, and forwarding-state consistency.

Table 1: Positioning of the present work against the closest Wi-Fi Direct multi-group studies.

| Study | Android mechanism | Core software context | Rooting | Main technical emphasis | Relevance to this paper |
|---|---|---|---|---|---|
| Di Felice *et al.* [3] | Smartphone testbed | Relay switches group affiliation over time | No | Emergency dissemination over multi-group smartphone networks | Establishes feasibility, but relies on relay switching and buffering by design. |
| Funai *et al.* [4] | Android 4.4.2 | Gateway node uses time sharing, multicast, or simultaneous legacy/WFD connectivity | No (stock) | Gateway behavior, delivery efficiency, and switching trade-offs | Shows that simultaneous connectivity is preferable to time sharing. |
| Casetti *et al.* [5] | Nonrooted Android | Application layer tunnels and contentcentric forwarding over a logical backbone | No | Contentcentric routing and trafficdirection effects | Identifies GO-originated broadcast as a major bottleneck. |

| This work | Stock Android 11 | Flutter application No layer, Kotlin native networking, relay-centered state coordination, and interface-bound sockets | Software architecture for multi-group communication under stockAndroid constraints | Demonstrates architectural feasibility of a stock-device artifact that preserves topicbased forwarding consistency during multi-group communication. |

3. Design requirements under stock Android constraints

The architecture was derived from constraints that are visible both in Android platform behavior and in prior WFD multi-group literature.

Table 2: Design requirements derived from the stock-Android setting.

| Requirement | Rationale |
| --- | --- |
| R1: Non-rooted operation | The solution must not rely on privileged routing-table changes or OS modification. |
| R2: Interfaceaware transport control | Inter-group forwarding requires distinct communication contexts for the WFD P2P and legacy-Wi-Fi interfaces. |
| R3: Explicit relay lifecycle management | Multi-group connectivity depends on relay promotion, demotion, and recovery as groups evolve. |
| R4: Forwardingstate consistency | Topic subscriptions must remain coherent as nodes join, subscribe, unsubscribe, or disconnect. |
| R5: Graceful fallback behavior | When relays fail or disappear, the group must continue operating in single-group mode rather than entering an inconsistent state. |
| R6: Practical implementability | The artifact must run on stock Android 11 devices using supported APIs and realistic deployment steps. |

R1 and R2 are driven directly by Android platform behavior. Android 6 and later changed how applications interact with network contexts, requiring developers to use modern multi-network APIs to ensure traffic is sent on a selected network [7, 8]. R3–R5 arise from the fact that multi-group communication is not a native platform service but an emergent behavior assembled at application level. Finally, R6 reflects the practical goal of building an artifact that works on stock devices rather than on modified prototypes.

## 4. SWARNET software architecture

### 4.1. Architectural overview

SWARNET is organized as a layered stock-Android artifact rather than as a monolithic routing implementation. Figure 1 and Figure 2 show the twogroup and three-group communication backbones realized by the artifact.

At software level, the architecture separates four concerns: application interaction, native network control, relay coordination, and subscription-based forwarding state.

Table 3: Main architectural elements and responsibilities.

| Element | Responsibility |
| --- | --- |
| Flutter application layer | User interaction, topic publish/subscribe actions, QR-code based bootstrap workflow, and high-level event orchestration. |
| Kotlin native networking layer | Access to Android Wi-Fi Direct APIs, group creation and discovery, socket creation, and network-context binding. |
| Relay manager | Promotion of a peer to relay, maintenance of primary/secondary relay roles, and inter-group communication orchestration. |
| Subscription-state manager | Maintenance of local subscription sets and forwarding tables; recomputation after join, subscribe, unsubscribe, and disconnect events. |
| Transport adapters | Interface-specific send/receive operations over P2P and legacyWi-Fi sockets. |

The architecture is intentionally centered on relays. In single-group mode, the GO remains the forwarding anchor. Once inter-group communication is required, a relay becomes the primary forwarding anchor for that group. This design moves inter-group forwarding away from GO-originated broadcast and allows inter-group legs to be carried through unicast transmissions.

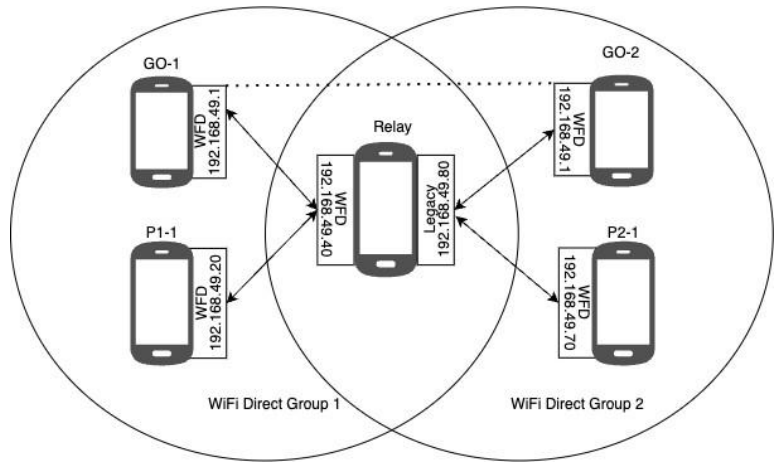

Figure 1: Implemented two-group communication backbone after relay promotion.

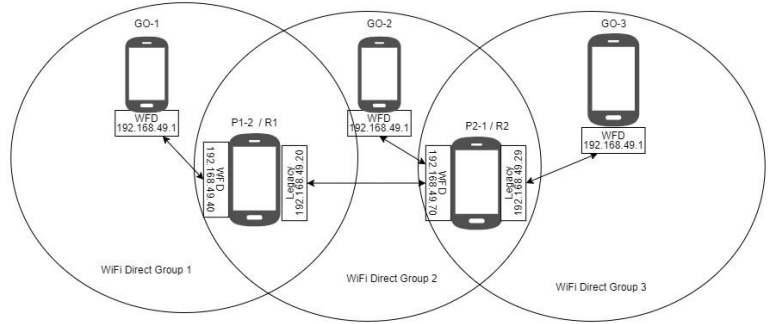

Figure 2: Implemented three-group communication backbone with primary and secondary relay roles.

*4.2. Interface-aware communication and socket binding*

A key architectural decision is the use of two communication contexts at a relay: one socket bound to the device's WFD P2P interface for nativegroup communication and one socket bound to the legacy-Wi-Fi connection for adjacent-group communication. This decision follows directly from the need to maintain simultaneous connectivity without rooting the device.

When a relay candidate joins an adjacent group as a legacy client, the application binds newly created sockets to the selected network context so that legacy traffic is kept on the intended interface. Existing WFD communication remains attached to the P2P socket context. In this way, the relay preserves connectivity to its native WFD group while also forwarding messages to the adjacent group.

This mechanism is central to the contribution because it transforms a platform restriction into an explicit architectural pattern: rather than trying to hide the existence of multiple interfaces, the software artifact models them directly and uses interface-bound transport adapters to control how traffic flows.

*4.3. Relay-state management*

The original manuscript mixed a theoretical relay-election formulation with an implemented procedure that selected an eligible peer and promoted it to relay. The present paper retains only the implemented behavior. A relay is an ordinary peer that is instructed to join an adjacent WFD group as a legacy client. If a group contains one relay, that relay acts as the primary relay (PR). If multiple relays are present, the relay native to the group acts as the PR and the remaining relay acts as a secondary relay (SR).

This distinction is important architecturally because it scopes responsibility. The PR may communicate with all forwarding endpoints in both its native and adjacent groups, while an SR communicates only with nodes in its native group and with the PR of its non-native group. The role split reduces ambiguity in forwarding behavior and simplifies state maintenance.

*4.4. Subscription routing and state consistency*

The original manuscript used the term "SRT" both for local node subscriptions and for network forwarding information. To remove this ambiguity, the revised architecture distinguishes two structures for each node $u$:

- a local subscription set $L_u$, containing the topics directly subscribed by $u$;

- a forwarding table $F_u(T)$, returning the next-hop set for publications on topic $T$.

At ordinary peers, $F_u(T)$ usually contains either the GO or the PR. At GOs and relays, the table aggregates downstream interests and may therefore contain multiple next hops. This separation is more faithful to the implemented software behavior and makes consistency rules explicit.

Table 4 gives an illustrative two-group state after relay promotion. The example is not intended as a full dump of all implementation variables; rather, it shows the architectural principle that the PR becomes the forwarding anchor for non-local topics once inter-group communication is active.

Table 4: Illustrative forwarding state after relay promotion in the two-group topology.

| Node $u$ | $\mathcal{L}_u$ | Representative forwarding state |
|---|---|---|
| $GO_1$ | $\emptyset$ | $\mathcal{F}_{GO_1}(T_2) = \{PR\}$ |
| $P_{1,1}$ | $\{T_1, T_2\}$ | $\mathcal{F}_{P_{1,1}}(T_2) = \{PR\}$ |
| PR | $\emptyset$ | $\mathcal{F}_{PR}(T_1) = \{P_{1,1}\}, \mathcal{F}_{PR}(T_2) = \{P_{1,1}, GO_2, P_{2,1}\}$ |
| $GO_2$ | $\emptyset$ | $\mathcal{F}_{GO_2}(T_2) = \{PR\}$ |
| $P_{2,1}$ | $\emptyset$ | $\mathcal{F}_{P_{2,1}}(T_2) = \{PR\}$ |

The artifact maintains consistency by recalculating forwarding state when one of four events occurs: node join, subscription update, publication, or disconnect.

**Algorithm 1** Publication forwarding in SWARNET

Require: Publication $m$, topic $T$, node $u$, local set $L_u$, forwarding table $F_u$
1: if $T \in L_u$ then
2:     Deliver $m$ to the local application at $u$
3: end if
4: for all $v \in F_u(T)$ do
5:     Send publication $(m, T)$ from $u$ to next hop $v$
6: end for

*Node join..* A joining node sends its communication address, group SSID, and local subscription set to the GO. If the group is operating in single-group mode, the GO updates the forwarding state and remains the forwarding anchor. If a PR already exists, the joining node is redirected to the PR, which integrates the node into the multi-group forwarding state.

*Subscription update..* When a node subscribes to or unsubscribes from topic $T$, the update is propagated to the GO or PR currently acting as forwarding anchor. The anchor recomputes the relevant forwarding entries and pushes updated routing information to affected nodes.

*Publication forwarding..* When a publication on topic $T$ is generated or received at node $u$, the node first checks whether $T \in L_u$ and delivers locally if necessary. It then forwards the publication to each next hop in $F_u(T)$.

*Disconnect handling..* Peers periodically send liveness beacons to the GO. When a membership change is detected, the GO classifies the event as either ordinary peer loss or relay loss. Peer loss triggers pruning and recomputation. Relay loss triggers either promotion of the remaining relay to PR or fallback to single-group GO-centered operation if no relay remains.

Algorithm 1 summarizes publication forwarding. Its forwarding complexity at node $u$ for topic $T$ is $O(|F_u(T)|)$.

*4.5. Implementation decisions and engineering trade-offs*

The architecture includes several pragmatic choices that should be understood as engineering decisions rather than as theoretically optimal mechanisms.

- Flutter plus Kotlin split. Flutter provides a portable application layer, while Kotlin retains direct access to Android networking APIs.
  This improves separation of concerns, although only the upper layer is potentially cross-platform.

- QR-based bootstrap. Adjacent-group credentials are currently transferred out of band using QR codes. This keeps the artifact compatible with stock devices but limits automation.

- Relay-centered forwarding. Once inter-group communication is enabled, the PR becomes the communication center. This removes dependence on GO-originated broadcast but also centralizes relay responsibility.

- Feasibility-oriented implementation. The current version prioritizes implementability on stock devices over advanced features such as

batteryaware relay selection, mobility adaptation, or predictive failure handling.

5. Artifact and empirical-validation methodology

*5.1. Artifact and platform*

The software artifact was deployed on five Samsung Galaxy A10s smartphones running Android 11 with vendor updates applied. The devices were factory reset before testing, and no third-party software other than the experimental application was installed. The application was implemented in Flutter and interfaced with a Kotlin native module responsible for Wi-Fi Direct control and socket-level networking.

Programming and deployment were carried out through IntelliJ IDEA with the Android development tools. All measurements were taken indoors with the devices placed in close proximity to minimize channel variations caused by distance and to make the experiments primarily reflect the behavior of the software artifact.

*5.2. Scenarios and metrics*

Four scenarios were exercised, progressing from single-group communication to two-group and three-group forwarding chains. Table 5 summarizes the scenarios.

Table 5: Validation scenarios used to exercise the software artifact.

| Case | Configuration | Source → destination | Forwarding path |
|---|---|---|---|
| 2d1g | 2 devices, 1 group | $P_{1,1} \to GO_1$ | single-hop, intra-group |
| 3d1g | 3 devices, 1 group | $P_{1,1} \to P_{1,2}$ | $P_{1,1} \to GO_1 \to P_{1,2}$ |
| 4d2g | 4 devices, 2 groups | $P_{1,1} \to P_{2,1}$ | $P_{1,1} \to PR \to P_{2,1}$ |
| 5d3g | 5 devices, 3 groups | $P_{1,1} \to GO_3$ | $P_{1,1} \to SR/PR \to PR \to GO_3$ |

The validation reported in the original artifact consists of throughput and packet-loss measurements as functions of offered load. Each point corresponds to the mean over 100 runs. These measures are retained here not as proof of a new routing theory, but as implementation-level evidence that the architecture remains operational as the software stack moves from singlegroup to multi-group communication.

A limitation of the current study is that the available experimental record contains only throughput and packet-loss plots. No new latency, energy, mobility, or relay-failure experiments were available at the time of writing, so the claims of the paper are confined to feasibility and observed communication behavior in a small static testbed.

## 6. Results

### *6.1. Observed communication behavior*

Figures 3–10 present the measured curves for the four scenarios. Table 6 summarizes the main observations.

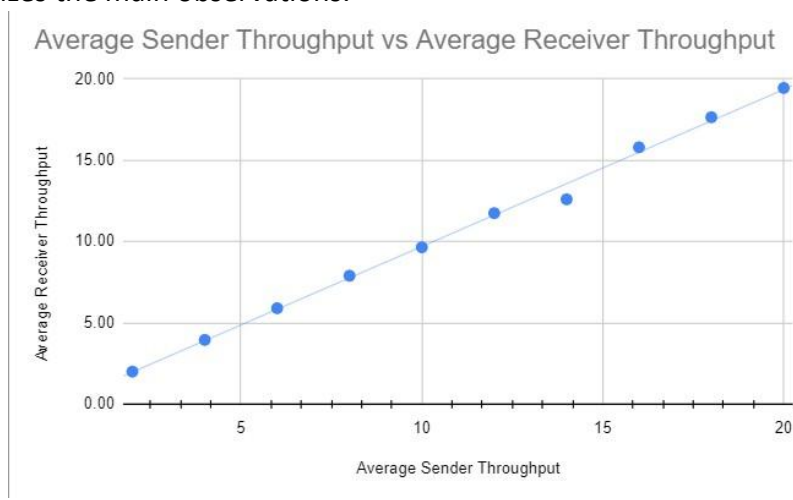

Figure 3: 2d1g: average sender throughput versus average receiver throughput.

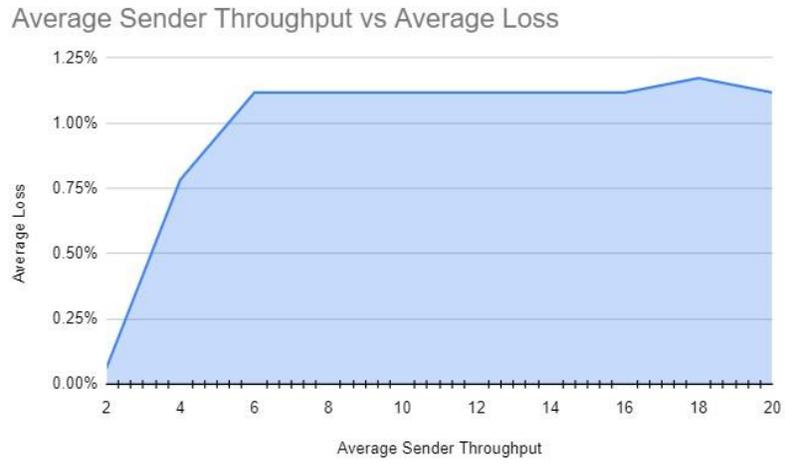

Figure 4: 2d1g: average sender throughput versus packet loss.

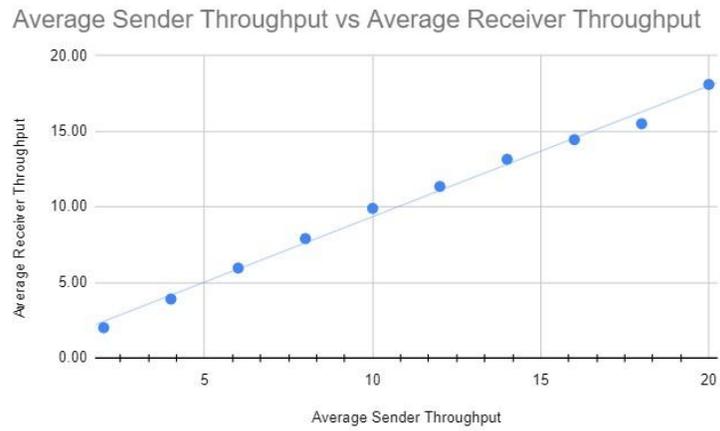

Figure 5: 3d1g: average sender throughput versus average receiver throughput.

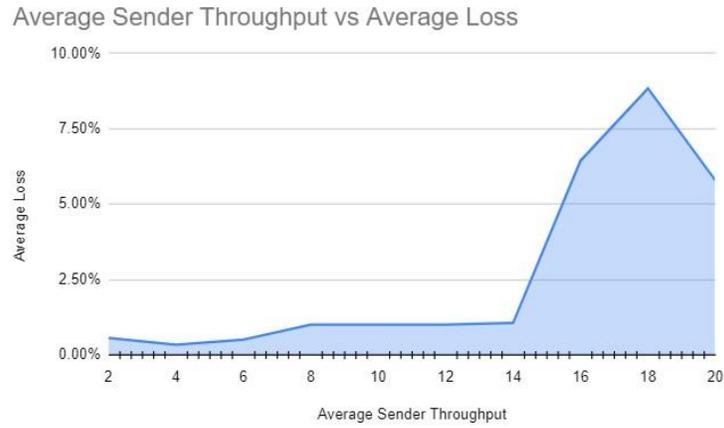

Figure 6: 3d1g: average sender throughput versus packet loss.

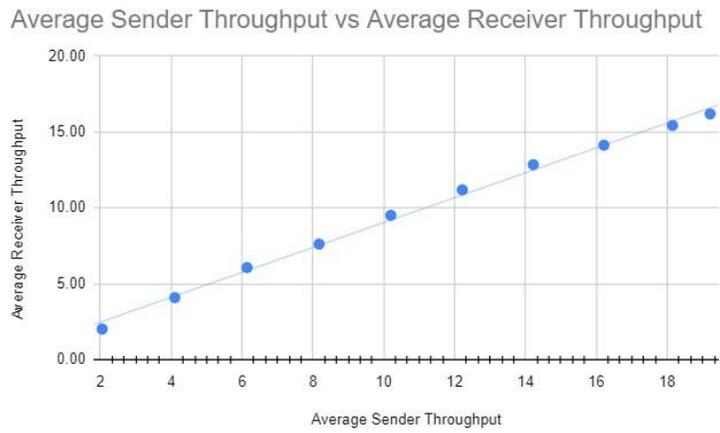

Figure 7: 4d2g: average sender throughput versus average receiver throughput.

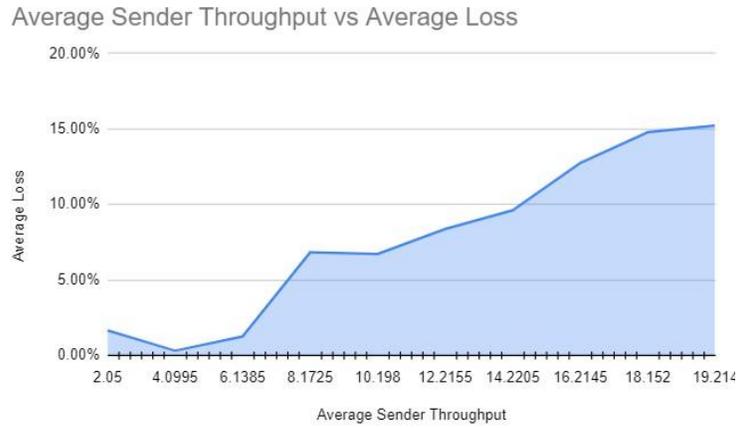

Figure 8: 4d2g: average sender throughput versus packet loss.

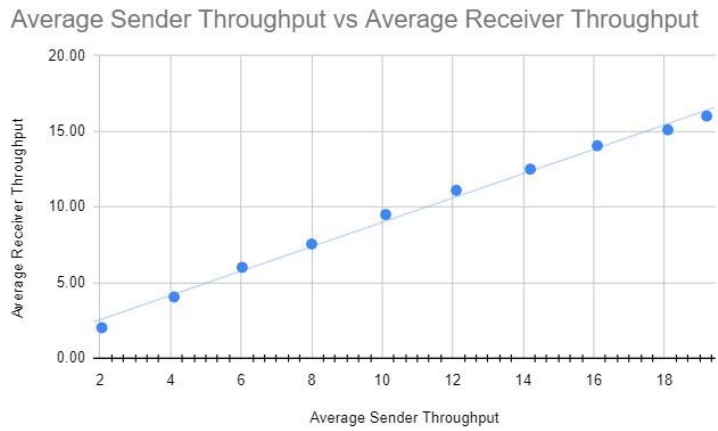

Figure 9: 5d3g: average sender throughput versus average receiver throughput.

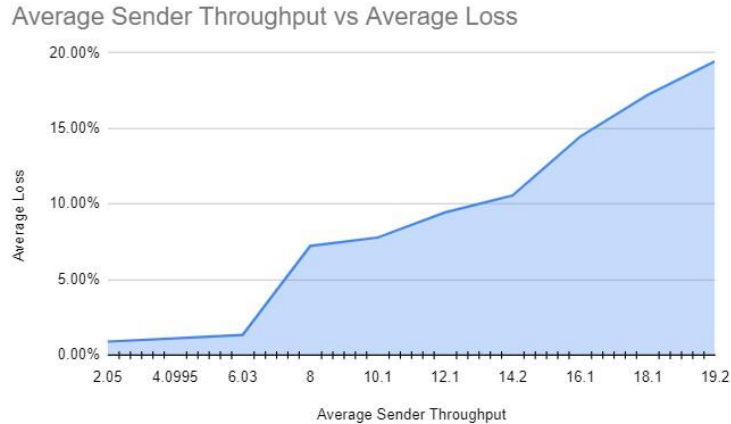

Figure 10: 5d3g: average sender throughput versus packet loss.

Table 6: Summary of observed behavior. Peak values are read from the reported curves.

| Case | Peak receiver throughput | Peak loss | Architectural interpretation |
|---|---|---|---|
| 2d1g | ≈ 19.7 Mbit/s | ≲ 1.25% | Baseline behavior of the artifact without relay processing |
| 3d1g | ≈ 17.9 Mbit/s | ≈ 9.5% | GO-mediated forwarding remains viable but saturates at high load |
| 4d2g | ≈ 16.1 Mbit/s | ≈ 15% | Two-group relay architecture remains operational under substantial offered load |
| 5d3g | ≈ 16.0 Mbit/s | ≈ 19–20% | Three-group chain remains feasible but becomes fragile in the highest-load regime |

The 2d1g case provides the implementation baseline. Receiver throughput scales almost linearly with sender throughput and approaches about 19.7 Mbit/s, while packet loss stays below approximately 1.25%. This indicates that the software stack imposes little visible penalty when no relay or GO-based forwarding stage is required.

The 3d1g case adds an application-level forwarding stage at the GO. Receiver throughput remains strong and approaches about 17.9 Mbit/s. Loss remains low through the lower and middle load range and then rises sharply near the highest offered loads, reaching roughly 9.5%. Architecturally, this suggests that the application remains stable under GO-mediated forwarding but that the forwarding anchor experiences increasing contention and buffering pressure at the upper end of the tested range.

The 4d2g case is the key validation scenario for the multi-group artifact. Receiver throughput continues to scale and reaches about 16.1 Mbit/s, while packet loss increases gradually toward approximately 15% only near the highest offered load. The significance of this result is architectural rather than purely numeric: it indicates that relay-centered interface binding and forwarding-state management remain workable when the application spans two groups.

The 5d3g case stresses the artifact with another forwarding stage. Peak receiver throughput remains close to the two-group value at about 16.0 Mbit/s, indicating that the architecture remained operational when an additional group was introduced under the tested conditions. The loss curve, however, rises toward approximately 19–20% in the highest-load regime. This supports architectural feasibility across three groups in the small static testbed, but it should not be interpreted as evidence of broad scalability.

*6.2. Contextualization against earlier work*

Table 7 provides a cautious cross-study comparison. Casetti *et al.* reported maximum application-layer throughputs of about 19 Mbit/s, 8.4 Mbit/s, and 5.0 Mbit/s for 2d1g, 3d1g, and 4d2g unicast scenarios, with substantially lower values when the forwarding path involved GO-originated broadcast [5]. Funai *et al.* likewise showed the value of simultaneous connectivity over timesharing approaches [4]. These comparisons are not same-hardware baselines, but they still help contextualize why a relay-centered stock-Android architecture is worthwhile.

Table 7: Representative contextual comparison with prior WFD multi-group studies.

| Study | Reported outcome | Relevance for the present architecture |
|---|---|---|
| Casetti *et al.* [5] | Maximum throughput approximately 19 Mbit/s (2d1g), 8.4 Mbit/s (3d1g), 5.0 Mbit/s (4d2g), with lower values when broadcast is involved | Supports the decision to avoid GOoriginated broadcast as the center of multi-group forwarding. |

| Funai et al. [4] | Simultaneous connectivity outperforms time sharing; hybrid relays transfer 10 MB in roughly 4–5 s in favorable conditions | Supports the use of simultaneous P2P and legacy-Wi-Fi contexts at relay nodes. |
|---|---|---|
| This work | Peak receiver throughput approximately 19.7 Mbit/s (2d1g), 17.9 Mbit/s (3d1g), 16.1 Mbit/s (4d2g), and 16.0 Mbit/s (5d3g) | Indicates that the implemented stock-Android artifact remains operational under progressively more complex multi-group configurations. |

7. Discussion

*7.1. What the artifact demonstrates*

The empirical results support a specific and limited conclusion: a stockAndroid software architecture can coordinate interface-aware relay communication, maintain topic-based forwarding state, and exhibit multi-group communication behavior in the tested scenarios without rooting or operatingsystem modification. The results do *not* establish a new general routing optimum, nor do they demonstrate resilience or scalability beyond this small static testbed.

This distinction is important for interpreting the contribution. In the current form, the artifact demonstrates that the difficult part of modern WFD multi-group support lies in software organization under platform constraints. Relay-state management, forwarding-table coherence, and interface-specific transport control are not peripheral details; they are the core mechanisms that make the multi-group behavior possible on stock devices.

*7.2. Portability and maintainability implications*

The architecture has partial portability. The Flutter layer suggests a path toward interface and application-logic reuse across platforms, but the networking core remains Android-specific because it depends on Android Wi-Fi P2P and network-binding APIs. Therefore, portability should be understood as an architectural intent and a consequence of separation of concerns, not as an empirically validated result.

Maintainability is also best discussed cautiously. The explicit decomposition into application logic, native networking, relay management, and subscription-state management should improve code organization and future modification. Likewise, the clearer distinction between local subscriptions and forwarding state reduces ambiguity in the implementation model. However, the study does not contain code metrics, developer studies, or longitudinal maintenance data. The maintainability claim is thus analytical rather than measured.

*7.3. Threats to validity*

Several threats limit the strength of the conclusions.

- Small homogeneous testbed. The validation used five identical Samsung Galaxy A10s smartphones. Behavior may differ on other vendors, chipsets, or Android customizations.

- Indoor static setting. All devices were placed in close proximity. The results therefore do not capture mobility, interference-rich environments, or spatial separation.
- Limited metrics. Only throughput and packet loss are reported from the available experimental record. No latency, control overhead, energy consumption, or relay-recovery timing data were collected.

- Semi-manual deployment. Adjacent-group provisioning currently depends on QR-code credential transfer, which constrains the level of automation demonstrated by the artifact.

- Cross-study comparison only. The contextual performance comparison with the literature does not use identical hardware or traffic generators and should therefore be read qualitatively.

*7.4. Implications for future work*

The next stage of this work should strengthen the artifact as a softwareengineering contribution rather than merely add more throughput curves. The most valuable extensions would be automated credential bootstrap, battery-aware or link-aware relay selection, controlled latency and energy instrumentation, relay-failure and mobility handling, and broader compatibility tests across Android devices and vendor implementations.

8. Conclusion

This paper reframed Wi-Fi Direct multi-group support on stock Android as a software-architecture problem and reported the design and empirical validation of SWARNET, a stock-Android artifact for multi-group communication. The architecture combines a Flutter application layer, a Kotlin native networking layer, interface-aware socket binding, relay-state coordination, and subscription-based forwarding consistency. The empirical study on Android 11 devices shows that the artifact remains operational across single-group, two-group, and three-group scenarios, with measured peak receiver throughputs of approximately 19.7, 17.9, 16.1, and 16.0 Mbit/s, respectively. These results support architectural feasibility on stock Android, while also making clear that the current artifact remains a small-scale proof of implementability rather than evidence of broad resilience, scalability, or deployment readiness.

Reproducibility and artifact availability

This manuscript is based on a preserved description of the original Android artifact, topology figures, and archived throughput/loss plots from the original study. The original Android source code and raw measurement logs were not recoverable as complete archival artifacts at the time this manuscript was prepared and are therefore not included with the submission. The materials available for inspection are the manuscript source, the figure files, and the summary tables derived from the preserved plots. Accordingly, the paper should be interpreted as an archival report of architectural design and preserved evaluation outputs rather than as a fully repeatable artifact package.